\documentclass[twocolumn,showpacs,pra]{revtex4} \usepackage{graphicx}
\usepackage{dcolumn} \usepackage{bm} \usepackage{amsmath}
\usepackage{amsfonts} \usepackage{amssymb}
\begin{document}
\title{Boson-Fermion Resonance Model in One Dimension}

\author{A. Recati$^{1,2}$, J.N. Fuchs$^{1,3}$, and W. Zwerger$^{1}$}

\affiliation{$^{1}$Institute for Theoretical Physics, Universit\"at Innsbruck, 
Technikerstrasse 25, A-6020 Innsbruck, Austria\\
$^{2}$ CRS BEC-INFM, Povo and ECT$^\star$, Villazzano, I-38050 Trento, Italy \\
$^{3}$Laboratoire de Physique des Solides, Universit\'e Paris-Sud, 
B\^atiment 510, F-91405 Orsay, France}

\date{\today}

\begin{abstract} 
We discuss the BCS-BEC crossover for one-dimensional spin $1/2$
fermions at zero temperature using the Boson-Fermion resonance model
in one dimension. We show that in the limit of a broad resonance, this model 
is equivalent to an exactly solvable single channel model, the so-called
modified Gaudin-Yang model. 
We argue that the one-dimensional crossover may be
realized either via the combination of a Feshbach resonance and a
confinement induced resonance or using direct photo-association
in a two-component Fermi gas with effectively one-dimensional dynamics. 
In both cases, the system may be driven 
from a BCS-like state through a molecular Tonks-Girardeau gas close to
resonance to a weakly interacting Bose gas of dimers.
\end{abstract}


\maketitle

\section{Introduction}

In this article, we consider the problem of attractive fermions in one
dimension (1D), having in mind current experiments on ultra-cold
two-component Fermi gases of atoms \cite{Regal,crossoverexp}. In these
systems, the $s$-wave interaction between fermions in different
internal states can be tuned using a Feshbach resonance. By changing
the interaction from weakly attractive to weakly repulsive via a
resonance where the interaction diverges, one can explore the
crossover from a BCS superfluid, when the attraction is weak and
pairing only appears in momentum space, to a Bose-Einstein condensate
(BEC) of molecular dimers \cite{crossoverth}.  Experiments are
currently investigating gases that are in a three dimensional regime
(3D). A different situation occurs if the gas is confined in a very
anisotropic cigar-shaped trap, like, e.g., in an atomic wire created
with optical lattices \cite{Esslinger} or on an atom chip
\cite{Reichel}. If the transverse confinement is strong enough, the
system effectively becomes 1D, i.e., the radial degrees of freedom are
frozen. We will refer to such a situation as quasi-1D.  In this case,
at zero temperature, a crossover takes place between a BCS-like state
and a weakly interacting Bose gas of dimers. This crossover can be
described by an exactly solvable model (the so-called modified
Gaudin-Yang model) \cite{FRZ,Tokatly}, which is just a combination of
the Gaudin-Yang model for attractive fermions \cite{GY} and of the
Lieb-Liniger model for repulsive dimers \cite{LL}. Despite the fact,
that there is no genuine off-diagonal long range order in 1D even at zero
temperature, we refer
to this situation as a one dimensional version of the BCS-BEC
crossover.

Such a crossover can be realized in two rather different ways using 
a two-component Fermi gas in a quasi-1D situation.
They correspond to 

\begin{itemize}

\item[i)]

fermions whose 3D-scattering length exhibits a Feshbach resonance (FBR).
In this case the combination of the 3D FBR and the confinement
in the transverse direction, charcterized by a trapping frequency 
$\omega_{\perp}/2\pi$, leads to a confinement induced (CI) resonance  
\cite{Olshanii}, beyond which
the two particle bound state energy is large enough to neglect breaking of
dimers (this scenario has been discussed in \cite{FRZ,Tokatly}).

\item[ii)]

Fermions which are transferred directly
into a bound molecular state by an external laser field.
The photo-association process can be described by an
effective 1D Boson-Fermion resonance model (BFRM) \cite{FL,RRE,RM}. For
positive detuning of the laser this describes a system of attractively
interacting fermions while
for negative detuning, one obtains again unbreakable dimers for strog enough
laser coupling.

\end{itemize}

The purpose of the
present paper is to study the 1D BFRM, ii), at zero temperature.
It will be shown that the resonance, which is reached by quite different
means in both cases, quite generally
allows driving a BCS-BEC crossover in 1D.  In
particular we will find
that in the BFRM the molecular size on resonance $r_{\star}$ plays a role 
similar to that of the transverse oscillator length $a_\perp\equiv
\sqrt{\hbar/m\omega_\perp}$ in the quasi-1D single channel model, i).
In the limit of low density $n$, characterized either by $na_\perp \ll 1$ \
or by $nr_{\star}\ll 1$ respectively, the resonance is broad 
and both models are completely equivalent to the exactly solvable
modified Gaudin-Yang model discussed in ref. \cite{FRZ,Tokatly}.  

The paper is organized as follows: in Sec. II we introduce the model
and the notations; Sec. III discusses the two-body problem, i.e.,
bound state and scattering properties; the many-body problem is
addressed in Sec. IV using a functional integral approach; and in
Sec. V we discuss the results.

\section{Boson-Fermion resonance model}

The Boson-Fermion resonance model \cite{FL,RM} is characterized by
the following (grand-canonical) Hamiltonian operator
\begin{eqnarray}
&\hat{H}'&=\hat{H}-\mu \hat{N}=\int dx \bigg(
\sum_{\sigma={\uparrow,\downarrow}} \hat{\psi}_{\sigma}^{\dagger}
\Big[-\frac{\hbar^2}{2m}\partial_x^2-\mu\Big]\hat{\psi}_{\sigma}
\nonumber \\
&+&\hat{\psi}_{B}^{\dagger}\Big[-\frac{\hbar^2}{4m}\partial_x^2-2\mu+\nu\Big]
\hat{\psi}_{B} + g \Big( \hat{\psi}_{B}^{\dagger}\hat{\psi}_{\uparrow}
\hat{\psi}_{\downarrow}+h.c.\Big)\bigg)\nonumber\\ \left.  \right.
\label{BFM}
\end{eqnarray}
where $\hat{\psi}_{\sigma}(x)$ (resp. $\hat{\psi}_B(x)$) are fermionic
(resp. bosonic) field operators describing atoms (resp. the bound
state in the closed channel, i.e., bare dimers), $\sigma$ identifies 
the spin projection $\uparrow$ or $\downarrow$,corresponding to the two
components in the Fermi gas, $\mu$ is a Lagrange
multiplier to be later identified with the chemical potential, $m$
(resp. $2m$) is the mass of the atoms (resp. of the bare dimers),
$\nu$ is the detuning in energy of one bare dimer with respect to two
atoms and $g$ is the coupling constant for the conversion of two atoms
into a bare dimer and vice-versa. 
The model Eq. (\ref{BFM}) can describe photo-association
of molecules in a 1D geometry. 
In this context, the coupling constant $g$ is
determined by the matrix element of the dipole energy, i.e., the
effective Rabi frequency,
and a Franck-Condon-factor, arising from the overlap of the 
wave functions of  atoms and  molecules. 
We assume the molecular size
to be much smaller than the oscillator length. 
Moreover we neglect the
background scattering between fermions, i.e., we do not include terms of the
form
$g_1^{bg}\hat{\psi}_{\uparrow}^{\dagger}
\hat{\psi}_{\downarrow}^{\dagger}\hat{\psi}_{\downarrow}
\hat{\psi}_{\uparrow}$ in the Hamiltonian. 
This is justified in any case close enough to resonance, i.e., where
$\nu\sim 0$ (see also Eq. (\ref{g1}).

The operator measuring the total number of atoms (i.e. unbound atoms 
and atoms bound into bare dimers) is:
\begin{eqnarray}
\hat{N}=\int dx \bigg( \sum_{\sigma={\uparrow,\downarrow}}
\hat{\psi}_{\sigma}^{\dagger}\hat{\psi}_{\sigma} +
2\hat{\psi}_{B}^{\dagger}\hat{\psi}_{B}\bigg).
\end{eqnarray}
We consider the zero temperature behavior of a system made of $N/2$
atoms with spin $\uparrow$ and $N/2$ atoms with spin $\downarrow$
confined on a ring of length $L$ and use the parameter $\mu$ to insure
that $\langle \hat{N} \rangle =N$. The thermodynamic limit is taken by
letting $N\to +\infty$ while maintaining the density $n\equiv N/L$
fixed. From now on, we set $\hbar=1$.

We note that the form of the local conversion term in 
Hamiltonian (\ref{BFM}) is fixed by the Pauli principle in a 
two-component  Fermi system. By contrast, for bosonic atoms, 
infinitely many local conversion terms 
$g_l (\hat{\psi}^{\dagger})^{l}\hat{\psi}_B^{(l)}$
with $l=2,3,4,..$ are possible and have to be considered. In order to
understand the relevance of such terms, including also a possible
background interaction between atoms, we performed a perturbative 
renormalization group (RG) analysis of the bosonized version of this model, i.e., an
atomic Bose Luttinger liquid converting into molecular Bose Luttinger
liquids. We find that if the background interaction is weak,
essentially all conversion terms are relevant. This implies that 
it is impossible to describe 1D bosonic atoms close to resonance in
terms of only a few parameters.

\section{Two-body problem: scattering amplitude and bound state}

In this section, we compute the molecular propagator in presence of
only two atoms ($N=2$).  From it, we obtain the scattering amplitude
between two atoms and the dressed (i.e.  renormalized) rest energy of
a dimer. The latter corresponds to the energy of a two-atoms bound
state. In momentum-energy space, the molecular propagator is given by
\begin{eqnarray}
D(k,\omega)=D_0(k,\omega)+D_0(k,\omega)\Pi(k,\omega)D(k,\omega)
\label{dysonvacuum}
\end{eqnarray}
where $D_0$ is the bare molecular propagator
\begin{eqnarray}
D_0(k,\omega)=\bigg[\omega-\frac{k^2}{4m}+2\mu-\nu+i0^+\bigg]^{-1}
\label{D0}
\end{eqnarray}
and the ``polarization'', i.e., self-energy of the closed channel
propagator, $\Pi(k,\omega)$ is given by:
\begin{eqnarray}
\Pi(k,\omega)=g^2 \int \frac{dk'}{2\pi}
\Big[\omega-k'^2/m-k^2/4m+2\mu+i0^+\Big]^{-1}.
\label{pi2}
\end{eqnarray}
From Eq. (\ref{dysonvacuum}), we can compute the dressed rest energy
$\epsilon_b$ of a dimer, which is defined as being the $k=0$ pole of
the molecular propagator when $\mu=0$:
\begin{eqnarray}
D(0,\epsilon_b)^{-1}=D_0(0,\epsilon_b)^{-1}-\Pi(0,\epsilon_b)=0.
\label{formalBS}
\end{eqnarray}
We find that Eq. (\ref{formalBS}) admits a unique real negative
solution $|\epsilon_b|=-\epsilon_b$ irrespective of the sign of 
the detuning $\nu$
\begin{equation}
\frac{|\epsilon_b|}{|\epsilon_{\star}|}-
\sqrt{\frac{|\epsilon_{\star}|}{|\epsilon_b|}}+
\frac{\nu}{|\epsilon_{\star}|}=0,
\label{bse}
\end{equation}
where we have introduced the on-resonance ($\nu=0$) bound state energy
$\epsilon_{\star}\equiv \epsilon_b(\nu=0)=-m^{1/3}g^{4/3}/2^{2/3}$.
This has to be compared with the 3D BFRM where a bound state is
present only when the detuning is negative \cite{footnote1}.
 
According to standard scattering theory, the $T$-matrix 
is given by $T=g^2D$ (see, e.g., \cite{DS}). Therefore, in the
BFRM, the Lippmann-Schwinger equation for atoms is equivalent to the
closed channel Dyson equation for the molecular propagator in vacuum,
equation (\ref{dysonvacuum}). From the latter it is possible to show
that the scattering between two atoms can be described as resulting
from an effective contact potential $g_1\delta(x)$, which is a well
defined 1D potential, with a bare scattering amplitude
\begin{eqnarray}
g_1\equiv g^2 D_0(0,0)=-\frac{g^2}{\nu}.
\label{g1}
\end{eqnarray}
When the detuning goes to zero, the bare scattering amplitude
diverges: this corresponds to the resonance. Before resonance, we have 
$\nu>0$ and an attractive effective interaction $g_1<0$ between the 
atoms, while after resonance $\nu<0$ and the effective interaction is 
repulsive $g_1>0$. Solving the Lippmann-Schwinger equation for the 
on-shell $T$-matrix of the contact potential
\begin{eqnarray}
T(k',k,\Omega)&=&g_1+i\int \frac{dk''}{2\pi}\frac{d\omega''}{2\pi} g_1
T(k',k'',\Omega)\nonumber \\ &\times&
\Big[(\Omega/2+\omega''-k''^2/2m+i0^+)\nonumber \\ &\times&
(\Omega/2-\omega''-k''^2/2m+i0^+)\Big]^{-1}
\end{eqnarray}
in the limit $k'=k \to 0$ and $\Omega=k^2/m$, we obtain the low-energy
limit of the one-dimensional two-body $T$-matrix:
\begin{eqnarray}
T_k=g^2D(0,k^2/m)\simeq \frac{g_1}{1+img_1/2k}.
\end{eqnarray}
The associated dressed scattering amplitude
\begin{eqnarray}
f(k)=\frac{m}{2ik}T_k \simeq -\frac{1}{1+ika_1}
\end{eqnarray}
has the standard form for 1D low energy scattering with 
$a_1\equiv -2/mg_1$ the 1D scattering length.  It is a
well-known fact that the 1D delta-potential forbids transmission at
low scattering energy, i.e., $f(k) \to -1$ in the $k\to 0$ limit
\cite{Olshanii}.

Before studying the many-body problem, we would like to discuss
briefly the behavior of the bound state energy.  We define the size of
the bound state as $r_b\equiv (m|\epsilon_b|/2)^{-1/2}$, which is
finite for any detuning, and call $r_{\star}\equiv r_b(\nu=0)$ 
the size of the bound state on resonance.  We find it useful also to
define a dimensionless detuning $\delta\equiv
\nu/\epsilon_{\star}\sqrt{2}$. With these definitions, equation
(\ref{g1}) becomes $g_1=2/mr_{\star}\delta$.

In the BCS limit (i.e. when $\delta \to -\infty$), the bound state
energy can be written as
\begin{eqnarray}
\epsilon_b\simeq -mg^2/4\nu^2=-mg_1^2/4
\end{eqnarray}
which agrees with the bound state energy of the $g_1\delta(x)$
potential when $g_1<0$.  In the opposite limit (BEC limit, i.e. when
$\delta \to +\infty$), the bound state energy is equal to the
detuning
\begin{eqnarray}
\epsilon_b\simeq \nu
\end{eqnarray}
and thus completely independent of the coupling constant $g$.
\begin{figure}
\begin{center}
\includegraphics[height=6cm]{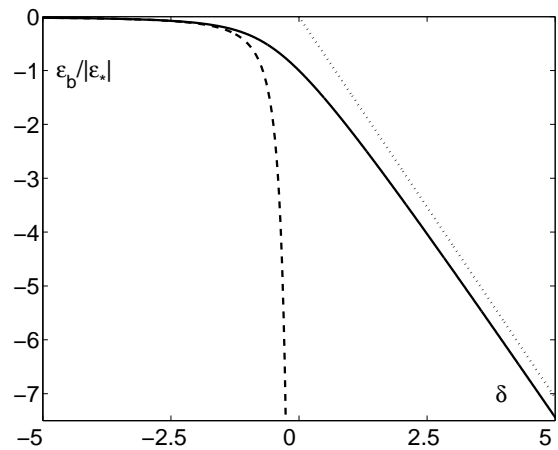}
\label{BFRMboundstate}
\caption{Two-body bound state energy $\epsilon_b$ [in units of the bound state energy 
on resonance $|\epsilon_{\star}|$] as a function of the dimensionless
detuning $\delta$ (full line).  The dashed line corresponds to the
asymptotic behavior $\epsilon_b \simeq -mg_1^2/4$ and the dotted line
to the asymptotic behavior $\epsilon_b \simeq \nu$.}
\end{center}
\end{figure}
The bound state energy $\epsilon_b$ is plotted as a function of the
dimensionless detuning $\delta$ in
Figure 1.

The behavior of the bound state in the 1D BFRM is qualitatively
similar to that of the confinement induced bound state found by
Bergeman, Moore and Olshanii \cite{BMO} for two atoms trapped in
a quasi-1D geometry (i.e., a waveguide with radial frequency
$\omega_{\perp}/2\pi$). This fact reveals the connection, at the
two-body level, between the 1D BFRM and the quasi-1D single channel
model. In Figure 2 we have plotted the confinement induced (CI) bound state as a function
of $\delta'$ (see below). In the quasi-1D case, the role of the dimensionless
detuning $\delta$ is played by the parameter $\delta'\equiv a_{\perp}/a-A$, where
$a_{\perp}\equiv (m\omega_{\perp})^{-1/2}$ is the radial oscillator
length, $a$ is the 3D scattering length and $A\equiv
-\zeta(1/2,1)/\sqrt{2}\simeq 1.0326$ \cite{footnote2}. In the
quasi-1D geometry, the 1D scattering amplitude shows a CI 
resonance \cite{Olshanii} and is given by
\begin{eqnarray}
g_1'\equiv 2\omega_{\perp}a(1-Aa/a_{\perp})^{-1}=2/ma_{\perp}\delta'
\end{eqnarray}
which is similar to $g_1=2/mr_{\star} \delta$, showing that
$a_{\perp}$ plays the role of $r_{\star}$. 
The CI bound state energy $\epsilon_b'$ obeys the following equation
\begin{eqnarray}
\sqrt{2}a_{\perp}/a+\zeta(1/2,\epsilon_b'/\epsilon_{\star}')=0
\end{eqnarray}
where $\zeta(1/2,x)$ is a particular Hurwitz zeta function \cite{BMO}
and $\epsilon_{\star}'\equiv \epsilon_b'(\delta'=0)=-2/ma_{\perp}^2$
is the CI bound state energy on resonance.  When $\delta' \to
-\infty$, $\epsilon_b'\simeq -mg_1'^2/4$, in complete analogy with 
$\epsilon_b\simeq -mg_1^2/4$. On resonance $\delta'=0$,
$\epsilon_{\star}'=-2/ma_{\perp}^2$, to be compared with
$\epsilon_{\star}=-2/mr_{\star}^2$. After resonance 
when $\delta' \to +\infty$, the CI bound state energy behaves as
$\epsilon_b'\simeq -1/ma^2$, which translates into 
$\epsilon_b'/\epsilon_{\star}'\simeq (\delta'+A)^2/2\simeq
\delta^{'2}/2$ in terms of the parameter $\delta'$ introduced 
above. Similarly, in the BFRM, the bound state energy decreases 
monotonically with the behavior $\epsilon_b/\epsilon_{\star} \simeq
\sqrt{2}\delta$ as a function of the dimensionless detuning 
$\delta$. 
\begin{figure}
\begin{center}
\includegraphics[height=6cm]{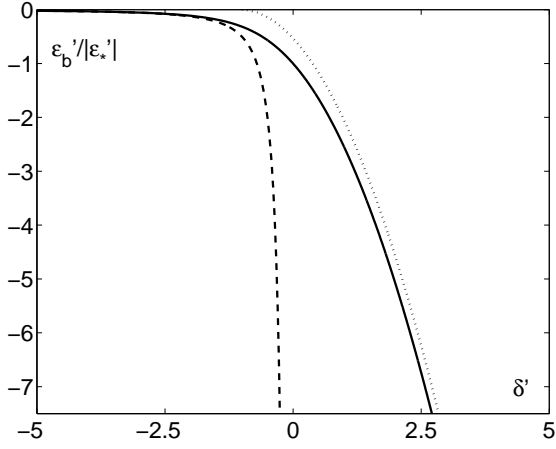}
\label{BMOboundstate}
\caption{Confinement induced bound state energy $\epsilon_b'$ [in units of the bound state energy 
on resonance $|\epsilon_{\star}'|$] as a function of the dimensionless
parameter $\delta'$ (full line).  The dashed line corresponds to the
asymptotic behavior $\epsilon_b' \simeq -mg_1^{'2}/4$ and the dotted
line to the asymptotic behavior $\epsilon_b' \simeq -1/ma^2$.}
\end{center}
\end{figure}

\section{Many-body problem}

The grand partition function at temperature $T\equiv 1/\beta$ and
chemical potential $\mu$ can be written as a path integral
\begin{eqnarray}
Z=\int \mathcal{D}(\bar{\psi}_{\sigma},\psi_{\sigma})
\mathcal{D}(\bar{\psi}_B,\psi_B) e^{-S}
\label{Z}
\end{eqnarray}
over Grassmann fields $\bar{\psi}_{\sigma}$, $\psi_{\sigma}$ with
$\sigma=\uparrow,\downarrow$ and complex fields $\bar{\psi}_B$,
$\psi_B$ \cite{AS}. The action corresponding to the Hamiltonian 
(\ref{BFM}) is:
\begin{eqnarray}
&S&=\int_0^{\beta}d\tau \int dx \bigg(
\sum_{\sigma={\uparrow,\downarrow}} \bar{\psi}_{\sigma}
\Big[\partial_{\tau}-\frac{\partial_x^2}{2m}-\mu\Big]\psi_{\sigma}
\nonumber \\
&+&\bar{\psi}_{B}\Big[\partial_{\tau}-\frac{\partial_x^2}{4m}-2\mu+\nu\Big]
\psi_{B} + g \Big( \bar{\psi}_{B}\psi_{\uparrow}
\psi_{\downarrow}+c.c.\Big)\bigg).\nonumber\\ \left. \right.
\label{originalaction}
\end{eqnarray}
The average total number of atoms is obtained from
\begin{eqnarray}
\langle N \rangle = \frac{\partial F}{\partial \mu}
\end{eqnarray}
where $F\equiv-T \ln Z$ is the grand potential and we are interested in
the $T\to 0$ limit.

Let us define the Fermi momentum $k_F\equiv \pi n/2$ and the Fermi
energy $\epsilon_F\equiv k_F^2/2m$ for an ideal gas of $N$ spin $1/2$
fermions. We shall study the particular case where the (modulus of
the) energy of the two body bound state on resonance is much larger
than the Fermi energy. This corresponds to the limit of a 
broad resonance (or strong coupling limit):
\begin{eqnarray}
|\epsilon_{\star}|\gg \epsilon_F \Leftrightarrow n r_{\star} \ll 1
\Leftrightarrow g\sqrt{n}\gg \epsilon_F
\label{BRLC}
\end{eqnarray}
The above inequality shows that the broad resonance limit corresponds to having a
deeply bound state after resonance, i.e., that the dimers are
unbreakable after resonance.  We will show that in this limit, the
system is described by a single channel model of atoms (fermions)
only, before resonance, and of dimers (bosons) only, after resonance.  
In other words, the 1D BFRM in the broad resonance 
limit is equivalent to the modified Gaudin-Yang model. 
We mention that equation (\ref{BRLC}) is similar to the usual criterion 
for a broad 3D Feshbach resonance \cite{BRL,footnote3}. 

\subsection{Before resonance: integrating out the bare dimers}
Before resonance $\nu>0$, it is possible to integrate out the bosonic
fields and describe the system in terms of an effective action for
fermions only. In order to show this, we need to define the Fourier
transform of a field $\psi(x,\tau)$:
\begin{eqnarray}
\psi(k,\tilde{\omega})&\equiv& \int d\tau dx
e^{i\tilde{\omega}\tau-ikx}\psi(x,\tau)\\ \psi(x,\tau)&\equiv& \int
\frac{dk}{2\pi} \frac{d\tilde{\omega}}{2\pi}
e^{-i\tilde{\omega}\tau+ikx}\psi(k,\tilde{\omega})\nonumber\\
&=&\int_{k,\tilde{\omega}}
e^{-i\tilde{\omega}\tau+ikx}\psi(k,\tilde{\omega})
\end{eqnarray}
When going to real time $t=-i\tau$, the analytic continuation on
frequencies is performed as $i\tilde{\omega}\to \omega +i0^+$.

Performing the Gaussian integration on $\bar{\psi}_B$ and $\psi_B$ in
equation (\ref{Z}) leads to:
\begin{eqnarray}
Z=Z_B^0 Z_F^{eff}=Z_B^0 \int
\mathcal{D}(\bar{\psi}_{\sigma},\psi_{\sigma}) e^{-S_F^{eff}}
\label{Z2}
\end{eqnarray}
where the effective action for fermions is
\begin{eqnarray}
S_F^{eff}&=&\int_{k,\tilde{\omega}}
\sum_{\sigma}\bar{\psi}_{\sigma}(k,\tilde{\omega})
\Big[-i\tilde{\omega}+\frac{k^2}{2m}-\mu\Big]\psi_{\sigma}(k,\tilde{\omega})\nonumber
\\ &+&g^2 \int_{k,\tilde{\omega}} \int_{k',\tilde{\omega}'}
\int_{K,\tilde{\Omega}}
\bar{\psi}_{\uparrow}(k,\tilde{\omega})\bar{\psi}_{\downarrow}(K-k,\tilde{\Omega}-\tilde{\omega})
\nonumber \\ &\times&
\psi_{\downarrow}(k',\tilde{\omega}')\psi_{\uparrow}(K-k',\tilde{\Omega}-\tilde{\omega}')
\nonumber\\ &\times& \Big[i\tilde{\Omega}-\frac{K^2}{4m}+2\mu-\nu
\Big]^{-1}
\end{eqnarray}
and the grand potential for an ideal Bose gas of bare dimers is:
\begin{eqnarray}
F_B^0&\equiv& -T\ln Z_B^0\nonumber \\ &=& TL \int \frac{dK}{2\pi} \ln
\left[1-e^{-\beta(K^2/4m-2\mu+\nu)}\right].
\end{eqnarray}

Due to the fact that only quadratic terms in $\psi_B$ appear in the 
original model, the previous result is exact. The resulting 
effective interaction between the atoms, however, is non-local both 
in space and time. If we restrict ourselves to the case 
$\nu >|\epsilon_{\star}|$, together with the broad resonance requirement
$|\epsilon_{\star}|\gg \epsilon_F$, we can simplify the effective
action $S_F^{eff}$ to one which is local. 
Indeed, before resonance, $2|\mu|\simeq
|\epsilon_b|<|\epsilon_{\star}|$ (see Appendix A) and as an order of
magnitude, $|i\tilde{\Omega}|\sim|K^2/4m|\sim \epsilon_F$.  Therefore,
the detuning dominates the denominator of the molecular propagator
$i\tilde{\Omega}-\frac{K^2}{4m}+2\mu-\nu \simeq -\nu$, and the
effective interaction between fermions becomes
\begin{eqnarray}
-\frac{g^2}{\nu} \int_{k_1,\tilde{\omega}_1}
\int_{k_2,\tilde{\omega}_2} \int_{k_3,\tilde{\omega}_3}
\bar{\psi}_{\uparrow}(1+2-3)
\bar{\psi}_{\downarrow}(3)\psi_{\downarrow}(2)\psi_{\uparrow}(1)\nonumber\\
\left. \right.
\end{eqnarray}
where $(1)$ is a short notation for $(k_1,\tilde{\omega}_1)$ 
and similarly for the other arguments. The total number of 
atoms can be computed from the partition function (\ref{Z2}):
\begin{eqnarray}
\langle N \rangle = -T\frac{\partial \ln Z_B^0}{\partial \mu}
-T\frac{\partial \ln Z_F^{eff}}{\partial \mu}.
\end{eqnarray}
The first term is given by the usual expression for an ideal Bose gas
\begin{eqnarray}
\langle N_B^0 \rangle=2L \int \frac{dK}{2\pi} \Big[ e^{\beta
\left(K^2/4m - (2\mu-\nu)\right)}-1\Big]^{-1}
\end{eqnarray}
with $2\mu-\nu<0$. When $T\to 0$, the fraction of atoms that are bound
into bare dimers is:
\begin{eqnarray}
\frac{\langle N_B^0 \rangle}{N}&\simeq& \frac{2}{n}
e^{-\beta(\nu-2\mu)} \int \frac{dK}{2\pi} e^{-\beta K^2/4m}\nonumber
\\ &=&\frac{2}{n}\sqrt{\frac{m}{\pi \beta}}e^{-\beta(\nu-2\mu)}\to 0.
\end{eqnarray}
Therefore
\begin{eqnarray}
N=\langle N \rangle \simeq -T\frac{\partial \ln Z_F^{eff}}{\partial
\mu}
\end{eqnarray}
which shows that $\mu$ is the chemical potential for the gas of atoms
only.

In conclusion, before resonance and under the assumptions that the
resonance is broad and that $\nu> |\epsilon_{\star}|$, the system is
described by a single channel model of fermions with an action
\begin{eqnarray}
S_F^{eff}&=&\int_0^{\beta}d\tau \int dx \bigg(
\sum_{\sigma={\uparrow,\downarrow}} \bar{\psi}_{\sigma}
\Big[\partial_{\tau}-\frac{\partial_x^2}{2m}-\mu_F\Big]\psi_{\sigma}
\nonumber \\ &+&g_1
\bar{\psi}_{\uparrow}\bar{\psi}_{\downarrow}\psi_{\downarrow}\psi_{\uparrow}
\bigg)
\end{eqnarray}
where $\mu_F=\mu$ and $g_1=-g^2/\nu<0$. This is the action
corresponding to the Gaudin-Yang model of 1D fermions interacting via
an attractive delta potential \cite{GY}.  The single dimensionless
coupling constant is $\gamma\equiv mg_1/n$. In order to describe the
BCS-BEC crossover, we will use the parameter $1/\gamma$ (see 
Appendix B), the BCS limit corresponding to $1/\gamma \to -\infty$ 
or $\nu \to +\infty$. Due to the
condition $\nu>|\epsilon_{\star}|$, before resonance, the parameter
$1/\gamma$ is restricted to:
\begin{eqnarray}
-\infty < \frac{1}{\gamma} < -\frac{n |\epsilon_{\star}|}{mg^2}\sim -n
 r_{\star}.
\end{eqnarray}
In the broad resonance limit, $nr_{\star} \to 0$ implying that apart
from a vanishingly small region close to resonance ($\nu=0$ or
$1/\gamma=0$), the Boson-Fermion resonance model, before resonance, is
equivalent to the single channel attractive Gaudin-Yang model.

\subsection{After resonance: integrating out the atoms}

After resonance ($\nu<0$), it is possible to integrate out the
fermionic fields and to describe the system in terms of an effective
action for dimers only. Formally, this is equivalent to the standard
technique used to study the single channel model in 2D or 3D (see
e.g. \cite{DZ, GT, Randeria93, PS}), where {\sl via} a
Hubbard-Stratonovich transformation, it is possible to write the
fermionic action in terms of a Bose-field only, which is eventually
identified with the order parameter of the superconducting phase. 
However, it is important to emphasize that the resulting bosonic field,
in that context 
is different from the field $\psi_B$ appearing in the BFRM defined by 
the action (\ref{originalaction}).

Performing the Gaussian integral over fermionic fields \cite{AS}, one
obtains
\begin{eqnarray}
Z=Z_F^0 Z_B^{eff}=Z_F^0 \int \mathcal{D}(\bar{\psi}_B, \psi_B)
e^{-S_B^{eff}}
\label{Z3}
\end{eqnarray}
where the effective action for bosons is
\begin{eqnarray}
S_B^{eff}&=&\int_{k,\tilde{\omega}} \bar{\psi}_{B}(k,\tilde{\omega})
\Big[-i\tilde{\omega}+\frac{k^2}{4m}-2\mu+\nu\Big]\psi_{B}(k,\tilde{\omega})
\nonumber \\ &-&\ln \det (G_0G^{-1})
\label{sbeff}
\end{eqnarray}
and $Z_F^0$ is the partition function for free fermions. We 
also defined the propagator for non interacting fermions, which 
in the Nambu representation reads
\begin{eqnarray}
G_0(k,\tilde{\omega};k',\tilde{\omega})
=\frac{(2\pi)^2\delta(k-k')\delta(\tilde{\omega}-\tilde{\omega}')}{
-i\tilde{\omega}+k^2/2m-\mu} \left(\begin{array}{cc} 1&0\\ 0&-1\\
\end{array}\right)
\end{eqnarray}
and the full propagator $G$ is related to $G_0$ by
\begin{eqnarray}
G^{-1}\equiv G_0^{-1}+\Delta
\end{eqnarray}
where $\Delta$ is given by:
\begin{eqnarray}
&&\Delta(k,\tilde{\omega};k',\tilde{\omega}) \equiv \nonumber\\
&-&g\left(\begin{array}{cc}
0&\psi_B(k+k',\tilde{\omega}+\tilde{\omega}')\\
\bar{\psi}_B(k+k',\tilde{\omega}+\tilde{\omega}')&0\\
\end{array}\right).
\end{eqnarray}
The last term in equation (\ref{sbeff}) can be expanded into a sum:
\begin{eqnarray}
-\ln \det(G_0G^{-1})
=\sum_{l=1}^{\infty}\frac{\text{Tr}\left[(G_0\Delta)^{2l}\right]}{2l}.
\label{suml}
\end{eqnarray} 
Below, we will evaluate explicitely the $l=1$ and $l=2$ terms in the sum
and, following \cite{GT,PS}, we will give an order of magnitude
estimate for the higher order terms showing that they are negligible in
the broad resonance limit.

\textbf{$l=1$ term}:
\begin{eqnarray}
\frac{1}{2}\text{Tr}\left[(G_0\Delta)^{2}\right]=
\int_{k,\tilde{\omega}}\bar{\psi}_B(k,\tilde{\omega})\psi_B(k,\tilde{\omega})
\Pi(k,\tilde{\omega})
\end{eqnarray} 
After resonance ($\nu<0$), we have $2|\mu|\simeq |\epsilon_b|>
|\epsilon_{\star}|\gg \epsilon_F$, as discussed in Appendix A. 
As an order of magnitude $|i\tilde{\omega}|$ gives a contribution 
of the order of the kinetic energy $\sim|k^2/4m|\sim \epsilon_F$. As a 
result, we obtain
\begin{eqnarray}
\Pi(k,\tilde{\omega})\simeq \Pi(0,0)\simeq \epsilon_b-\nu
\end{eqnarray}
where we used $\mu\simeq \epsilon_b/2$ and the bound state equation
(\ref{bse}). Thus the $l=1$ term just gives rise to an effective chemical 
potential for the dimers equal to $2\mu -\epsilon_b$.

\textbf{$l=2$ term}:
\begin{eqnarray}
\frac{1}{4}\text{Tr}\left[(G_0\Delta)^{4}\right]&=& \frac{1}{2}
\int_{k_1,\tilde{\omega}_1} \int_{k_2,\tilde{\omega}_2}
\int_{k_3,\tilde{\omega}_3} \bar{\psi}_{B}(1+2-3)\nonumber \\ &\times&
\bar{\psi}_{B}(3)\psi_{B}(2)\psi_{B}(1)g_B(1,2,3)
\end{eqnarray} 
in an obvious short hand notation for the associated wave vectors and
frequencies, and
\begin{eqnarray}
g_B(1,2,3)&\equiv& g^4\int_{k,\tilde{\omega}}\Big[
(i\tilde{\omega}-\xi_k) \nonumber \\ &\times&
(i\tilde{\omega}_2-i\tilde{\omega}-\xi_{k_2-k}) \nonumber \\ &\times&
(i\tilde{\omega}_1+i\tilde{\omega}_2-i\tilde{\omega}_3-i\tilde{\omega}-\xi_{k_1+k_2-k_3-k})
\nonumber \\ &\times&
(-i\tilde{\omega}_2+i\tilde{\omega}_3+i\tilde{\omega}-\xi_{-k_2+k_3+k})
\Big]^{-1}
\end{eqnarray}
with $\xi_k\equiv k^2/2m - \mu$. Using again the condition characterizing 
a broad resonance, we see that the momentum and frequency dependance of the 
interaction is irrelevant. This implies that
\begin{eqnarray}
g_B(1,2,3)&\simeq& g_B(k_j=0,\tilde{\omega}_j=0;j=1,2,3) \nonumber \\
&\simeq& g^4\int_{k,\tilde{\omega}} \Big[\tilde{\omega}^2+\xi_k^2
\Big]^{-2}=\frac{3g^4\sqrt{m}}{8|\epsilon_b|^{5/2}}
\label{gB}
\end{eqnarray}
with $\mu\simeq \epsilon_b/2$.

\textbf{$l\geq 3$ term}: For all $l$, we obtain the following estimate
for the corresponding term in the sum (\ref{suml}):
\begin{eqnarray}
t_l\equiv \frac{1}{2l}\text{Tr}\left[(G_0\Delta)^{2l}\right]\sim
\frac{g^{2l}}{|\epsilon_b|^{2l-1}r_b}\frac{k_F^{l-1}}{\epsilon_F}\sim
(nr_b)^{l-3}.
\end{eqnarray}
For $l=1$ and $l=2$, in the broad resonance limit where $1\gg
nr_{\star} >nr_b$, we obtain that $t_1$ and $t_2\gg 1$. For $l\geq 3$,
the ratio $t_2/t_l\gg 1$ in the broad resonance limit and the corresponding 
terms can therefore be neglected.

In the broad resonance limit, the effective action for
bosons becomes
\begin{eqnarray}
S_B^{eff}&=&\int_0^{\beta}d\tau \int dx \bigg(\bar{\psi}_{B}
\Big[\partial_{\tau}-\frac{\partial_x^2}{4m}-(2\mu-\epsilon_b)\Big]\psi_{B}
\nonumber \\ &+& \frac{g_B}{2}
\bar{\psi}_{B}\bar{\psi}_{B}\psi_{B}\psi_{B} \bigg)
\end{eqnarray}
where $g_B\equiv 3g^4\sqrt{m}/8|\epsilon_b|^{5/2}$ describes a repulsive 
interaction between the strongly bound dimers. From the
partition function (\ref{Z3}) and the preceding effective action, we
can obtain the average total number of atoms:
\begin{eqnarray}
\langle N \rangle = -T\frac{\partial \ln Z_F^0}{\partial \mu}
-T\frac{\partial \ln Z_B^{eff}}{\partial \mu}.
\label{N45}
\end{eqnarray}
The first term is given by the usual expression for an ideal Fermi
gas:
\begin{eqnarray}
\langle N_F^0 \rangle=L \int \frac{dk}{2\pi} \Big[ e^{\beta
\left(k^2/2m -\mu\right)}+1\Big]^{-1}.
\end{eqnarray}
In the limit $T\to 0$ and using the fact that after resonance 
$\mu\simeq \epsilon_b/2$, the fraction of atoms that are unbound is 
exponentially small. Therefore, (\ref{N45}) becomes
\begin{eqnarray}
\frac{N}{2}=\frac{\langle N \rangle}{2} \simeq -T\frac{\partial \ln
Z_B^{eff}}{\partial 2\mu}
\end{eqnarray}
which shows that $2\mu$ is the chemical potential for the gas of
dimers only. We now shift the zero of energy of the many-body 
system by an amount $-N\epsilon_b/2$, and accordingly define 
$\mu_B\equiv 2\mu -\epsilon_b$ as the new chemical potential 
for dimers.

In conclusion, after resonance and under the assumption that the
resonance is broad, the system is described by a single channel model
of bosons (i.e. dimers) with an action
\begin{eqnarray}
S_B^{eff}&=&\int_0^{\beta}d\tau \int dx \bigg(\bar{\psi}_{B}
\Big[\partial_{\tau}-\frac{\partial_x^2}{2m_B}-\mu_B\Big]\psi_{B}
\nonumber \\ &+&\frac{g_B}{2}
\bar{\psi}_{B}\bar{\psi}_{B}\psi_{B}\psi_{B} \bigg)
\end{eqnarray}
where $m_B\equiv 2m$, $\mu_B=2\mu-\epsilon_B$ and $g_B=
3g^4\sqrt{m}/8|\epsilon_b|^{5/2}$ \cite{footnote4}. This is the
action corresponding to the Lieb-Liniger model of $N_B\equiv N/2$ bosons of mass
$m_B$ interacting via a repulsive delta potential \cite{LL}. The single 
dimensionless coupling constant is $\gamma\equiv mg_B/n$ and 
the BEC limit corresponds to $1/\gamma \to +\infty$ 
or $\nu \to -\infty$.  Because
$g_B \sim 1/mr_{\star}$ when $\nu \to 0^{-}$ (see Appendix B), the 
parameter $1/\gamma$ is restricted to:
\begin{eqnarray}
n r_{\star} <\frac{1}{\gamma} < +\infty
\end{eqnarray}
In the broad resonance limit, $nr_{\star} \to 0$ implying that apart
from a vanishingly small region close to resonance the Boson-Fermion
resonance model, after resonance, is equivalent to the single channel
repulsive Lieb-Liniger model for dimers.

\section{Discussion}

It was recently shown \cite{FRZ,Tokatly} that interacting fermions in
a quasi-1D geometry and in presence of a Feshbach resonance map onto
the modified Gaudin-Yang model in the limit of very strong confinement
$na_{\perp}\ll 1$. In the present paper, we have seen that the
Boson-Fermion resonance model in 1D is also described by the same
model in the limit of a broad resonance $nr_{\star}\ll 1$.  Close to
resonance, the system behaves as a Tonks-Girardeau gas (or
impenetrable Bose gas) \cite{Girardeau} of dimers
\cite{Astra,FRZ,Tokatly}. Around resonance, there is a vanishingly
small region $1/\gamma \in [-nr_{\star}, nr_{\star} ]$, which is not 
described by the modified Gaudin-Yang
model. The relation between the parameters $g_1$ and $g_B$ 
of the modified Gaudin-Yang model and those of the original system -- 
either the quasi-1D single channel
model or the 1D Boson-Fermion resonance model -- is different. Deep in
the BEC limit, the two-body bound state of the BFRM is given by
$\epsilon_b\simeq \nu$ and is populated by pairs of fermions.  All
fermions are bound into dimers and the scattering properties of dimers
have no direct relation to the scattering properties of fermions, in
particular $g_B$ is not simply proportional to $g_1$. This is in
contrast to the 3D single channel model, where it is known that the
dimer-dimer scattering length is proportional to the fermion-fermion
scattering length (see, e.g., \cite{Petrov}). The equivalent result
for the quasi-1D single channel is still under investigation
\cite{Mora}. Nevertheless, in the BEC limit, one expects that $g_B
\approx 0.6 \, g_1$ \cite{FRZ,Tokatly}.

The above scenario for a BCS-BEC crossover can be
realized, e.g., in an experiment with ultra-cold gases confined in
a quasi-1D trap either by tuning the 3D $s$-wave scattering length via 
a magnetic field in order to cross the CI resonance, as discussed
in \cite{FRZ}, or by photoassociation \cite{photo}, which corresponds 
to a direct implementation of the BFRM. 
As mentioned before, a description of the
resulting 1D BCS-BEC crossover by the modified Gaudin-Yang model is possible 
under the condition (19) for a broad resonance. Specifically, the realization
using a CI resonance in a tight waveguide requires a sufficiently dilute gas
with $na_{\perp}\ll 1$. Taking typical values of order 50 nm for the
transverse oscillator length which have been realized very recently
in bosonic 1D gases \cite{Esslinger, BlochKinoshita}, this requires
densities in the range of much less than 20 atoms per micron.
In the case of photo-association, i.e. an optically induced resonance,
the requirement is, that the effective 1D Rabi frequency $g\sqrt{n}$ is
much larger than the Fermi energy. Using estimates for the Rabi-frequency
taken over from photassociation of $^{87}$Rb in 3D \cite{GrimmRb},
a rough estimate shows  that the condition of a broad resonance can
also be reached here. In particular  the fact that the Franck-Condon overlap
is enhanced in a 1D situation helps realizing this limit.

\bigskip

We acknowledge useful discussions with Walter Rantner, Stefano
Cerrito and Andrea Micheli. Laboratoire de Physique des Solides is a mixed research unit
(UMR 8502) of the CNRS and the Universit\'e Paris-Sud in Orsay.

\section*{Appendix A}
The estimates of $\mu$ used in the present article come from
identifying $\delta \mu\equiv \mu -\epsilon_b/2$ (when in the broad
resonance limit) with the chemical potential in the modified
Gaudin-Yang model \cite{FRZ}. The chemical potential obtained from
\cite{FRZ} gives the following estimate for $\delta \mu$:
\begin{displaymath}
\delta \mu/\epsilon_F \simeq \left\{ \begin{array}{ll} 1 & \text{ when
} 1/\gamma \to -\infty \text{ BCS limit}\\ 1/4 & \text{ when }
1/\gamma \to 0 \text{ on resonance}\\ \gamma/4\pi^2 & \text{ when }
1/\gamma \to +\infty \text{ BEC limit}
\end{array}\right.
\end{displaymath}

\section*{Appendix B}
In this appendix, we discuss the behavior of $1/\gamma$ as a function
of $\nu$. Before resonance $\gamma=mg_1/n=-mg^2/n\nu$, which implies:
\begin{eqnarray}
\frac{1}{\gamma}=-\frac{nr_{\star}}{2^{3/2}}\frac{\nu}{|\epsilon_{\star}|}.
\end{eqnarray}
In the BCS limit $\nu\to +\infty$, $1/\gamma \to -\infty$ and on 
resonance $\nu \to 0^{+}$, $1/\gamma \to 0^-$. We assumed that 
$|\nu|>|\epsilon_{\star}|$, which implies 
$1/|\gamma|> nr_{\star}/2^{3/2}$ with $nr_{\star}\ll 1$ (broad 
resonance limit), so that indeed, close to resonance 
$1/\gamma \simeq 0$.

After resonance, $\gamma=mg_B/n$ and equation (\ref{gB}) can be
rewritten \cite{footnote6}
\begin{eqnarray}
g_B=\frac{3}{\sqrt{2}mr_{\star}}\left(\frac{\epsilon_b}{\epsilon_{\star}}\right)^{-5/2}
\end{eqnarray}
which implies:
\begin{eqnarray}
\frac{1}{\gamma}=\frac{\sqrt{2}nr_{\star}}{3}
\left(\frac{\epsilon_b}{\epsilon_{\star}}\right)^{5/2}.
\end{eqnarray}
Close to resonance $\nu\to 0^{-}$, $g_B\simeq 3/\sqrt{2}mr_{\star}$
and $1/\gamma \simeq \sqrt{2}nr_{\star}/3$ with $nr_{\star}\ll 1$ 
(broad resonance limit), so that again, $1/\gamma \simeq 0$ close to
resonance. In the BEC limit $\nu \to -\infty$, $g_B \simeq
3\epsilon_{\star}^{5/2}/\sqrt{2}mr_{\star}\nu^{5/2}$ and
$1/\gamma\simeq \sqrt{2}nr_{\star}\nu^{5/2}/3\epsilon_b^{5/2} \to
+\infty$.


\begin{thebibliography}{99}
\bibitem{footnote1}In the 3D version of the model defined by the
Hamiltonian (\ref{BFM}), a renormalized detuning appears in the bound
state energy equation instead of the bare one as in the 1D case. This
is due to the fact that a delta interaction is not a valid 3D
potential and needs to be renormalized \cite{RM}.
\bibitem{footnote2}$\zeta(1/2,x)$ is a particular Hurvitz zeta
function and $\zeta(1/2,1)\simeq-1.4603$, see \cite{BMO}.  Note the
$\sqrt{2}$ difference in the definition of $a_{\perp}$ with
\cite{Olshanii,BMO}, which accounts for $A\equiv
-\zeta(1/2,1)/\sqrt{2}\simeq 1.0326$.
\bibitem{footnote3}Pistolesi and Strinati call it ``the bosonization
condition'' \cite{PS}. In 3D, this condition corresponds to requiring
that the effective range of the interaction between atoms
mediated by molecules is small, i.e. $n^{1/3}|r_{eff}|\ll 1$ with
$r_{eff}=-8\pi/m^2g^2$, see Bruun \cite{BRL} and S. De Palo et
al. \cite{BRL}.
\bibitem{footnote4}In the 3D resonance model, $g_B$ is given by the
same expression as in (\ref{gB}) except that the $k$ integral is over
three dimensions. This gives $g_B=m^{3/2}g^4/16\pi |\nu|^{3/2}$ in the
BEC limit, where $\epsilon_b\simeq \nu$. Using the previous result, to
calculate the Bogoliubov sound velocity $c\equiv \sqrt{g_Bn_B/m_B}$ 
with $n_B\equiv n/2$,
we recover the result of Ohashi and Griffin for the sound velocity in
the BEC limit of the 3D resonance model, see equation (4.20) in
\cite{OG}.
\bibitem{footnote6}We note that this is of the form
$(ma_{\perp})^{-1}f(|\epsilon_b|/2\omega_{\perp})$, with the
identification $r_{\star} \leftrightarrow a_{\perp}$ and
$|\epsilon_{\star}| \leftrightarrow 2\omega_{\perp}$, as discussed in
\cite{Tokatly}.
\end{thebibliography}
\end{document}